\documentclass[pre,twocolumn,superscriptaddress,floatfix]{revtex4}
\usepackage[latin1]{inputenc}
\usepackage{graphicx}
\usepackage{amsmath}
\usepackage{amssymb}
\usepackage{pifont}
\usepackage[dvips]{color}
\usepackage{psfrag}
\begin{document}
\title{Anisotropic flow in striped superhydrophobic channels}

\author{Jiajia Zhou}
\affiliation{Institut f\"ur Physik, Johannes Gutenberg-Universit\"at Mainz, 55099 Mainz, Germany}
\author{Aleksey V. Belyaev}
\affiliation{Department of Physics, M.V. Lomonosov Moscow State University, 119991 Moscow, Russia}
\affiliation{A.N. Frumkin Institute of Physical Chemistry and Electrochemistry, Russian Academy of Sciences, 31 Leninsky Prospect, 119991 Moscow, Russia}
\author{Friederike Schmid}
\affiliation{Institut f\"ur Physik, Johannes Gutenberg-Universit\"at Mainz, 55099 Mainz, Germany}
\author{Olga I. Vinogradova}
\affiliation{Department of Physics, M.V. Lomonosov Moscow State University, 119991 Moscow, Russia}
\affiliation{A.N. Frumkin Institute of Physical Chemistry and Electrochemistry, Russian Academy of Sciences, 31 Leninsky Prospect, 119991 Moscow, Russia}
\affiliation{DWI, RWTH Aachen, Forckenbeckstr. 50, 52056 Aachen, Germany}

\date{\today}

\begin{abstract}

We report results of dissipative particle
dynamics simulations  and develop a semi-analytical theory of an anisotropic flow in a parallel-plate channel with
two superhydrophobic striped walls. Our approach is valid for any local slip at
the gas sectors and an arbitrary distance between the plates, ranging from a
thick to a thin channel.  It allows us to optimize area fractions, slip
lengths, channel thickness and texture orientation to maximize a transverse
flow.  Our results may be useful for extracting effective slip tensors from
global measurements, such as the permeability of a channel, in experiments or
simulations, and may also find applications in passive microfluidic mixing.

\end{abstract}


\maketitle

\section{Introduction}

Textured surfaces play a major role in microfluidics, since the high surface to
volume ratio enhances fluid-surface interactions~\cite{stone2004}. An important
class of phenomena involve ``transverse'' hydrodynamic couplings in anisotropic
channels, where an applied pressure gradient or shear rate in one direction
generates flow in a different direction, with a nonzero perpendicular
component. Transverse hydrodynamic couplings in pressure-driven flow through a
textured microchannel were analyzed
theoretically~\cite{stroock2002b,ajdari2002,wang2003,bazant08} and applied to
passive chaotic mixing in a herringbone grooved
channel~\cite{stroock2002b,stroock2004}. Such microfluidic devices have also
recently been used to separate or concentrate suspended
particles~\cite{gao.c:2008}.

Here we deal with a special (and different) type of surface textures, namely a
superhydrophobic (SH) surface in the Cassie state, where air micro- and
nanobubbles are favored, and can generate a number of amazing
properties~\cite{quere.d:2005}, including a very large liquid
slippage~\cite{bocquet2007,vinogradova.oi:2010}.  Such a slip also occurs at
smooth hydrophobic surfaces~\cite{vinogradova1999,lauga2007}, but with a
relatively low amplitude, characterized by a slip length (extrapolated distance
on which the liquid velocity vanishes) of the orders of tens of nm or
smaller~\cite{vinogradova:03,charlaix.e:2005,joly.l:2006,vinogradova.oi:2009}.
Due to a local slip length of tens of $\mu$m over gas regions the dramatic
enhancement of slip at such a superhydrophobic surface could be
achieved~\cite{choi.ch:2006,joseph.p:2006,tsai.p:2009,rothstein.jp:2010} and may
dramatically reduce viscous drag in a channels.

\begin{figure}
\includegraphics[width=0.3\textwidth]{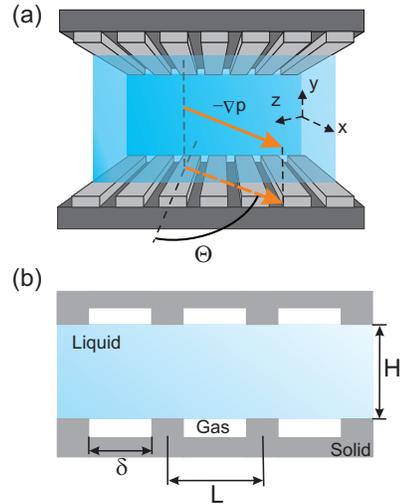}
\caption{Sketch of the symmetric striped channel (a): $\Theta=\pi/2$ corresponds to
 transverse stripes,
$\Theta=0$ to longitudinal stripes; (b) situation
in (a) is approximated by a periodic cell of size $L$, with equivalent flow
boundary conditions on the gas-liquid and solid-liquid interfaces.
}\label{fig:geometry}
\end{figure}

 Directional SH textures generate the anisotropy of effective  slip, which becomes a tensor,
$\textbf{b}_{\rm eff}\equiv\{b^{\rm eff}_{ij}\}$, represented by a symmetric,
positive definite $2\times 2$ matrix
\begin{equation} \label{beff_def1}
\mathbf{b}_{\rm eff} = \mathbf{S}_{\Theta}\left(
  \begin{array}{cc}
    b^{\parallel}_{\rm eff} & 0 \\
    0 & b^{\perp}_{\rm eff} \\
  \end{array}\right) \mathbf{S}_{-\Theta},
\end{equation}
diagonalized by a rotation with angle $\Theta$
\begin{equation} \label{S_def1}
\mathbf{S}_{\Theta}=\left(
  \begin{array}{cc}
\cos \Theta & \sin \Theta \\
- \sin \Theta & \cos \Theta \\
  \end{array}
\right).
\end{equation}

The tensor formalism allows us to easily change the orientation of a texture,
once a problem has been solved for a given geometry. For all anisotropic
surfaces its eigenvalues $b^{\parallel}_{\rm eff}$ and $b^{\perp}_{\rm eff}$
correspond to the fastest (greatest forward slip) and slowest (least forward
slip) orthogonal directions~\cite{bazant08}.  Note that the concept of an
effective slip length tensor is applied for an \emph{arbitrary} channel
thickness~\cite{harting.j:2012}, being a global characteristic of a
channel~\cite{vinogradova.oi:2010}. A corollary of this is  that the
eigenvalues depend not only on the parameters of the heterogeneous surfaces
(such as local slip lengths, fractions of phases, and a texture period), but
also on the channel thickness. However for a \emph{thick} (compared to a
texture correlation length) channel they become a characteristics of a
heterogeneous interface solely~\cite{bazant08}. In the general case of any
direction $\Theta$, the anisotropy of effective slip means that the flow past
such surfaces becomes misaligned with the driving force. Therefore, anisotropic
SH textures can potentially be used to generate transverse hydrodynamic flow,
which is of obvious fundamental and practical interest. However, most of the
prior theoretical work has focused on the maximization of the effective slip, and general
principles to optimize transverse phenomena have received much less attention
and have not yet been established.

Computer simulations might be expected to shed some lights on a flow past anisotropic surfaces. Earlier Molecular Dynamics (MD) simulations observed how modulations of hydrophobicity induce (small) variations in  local slip at channels of finite thickness, but did not attempt to relate these to effective boundary conditions~\cite{qian2005,
hendy2005}. In efforts to better understand the connection between the parameters of the texture and the effective slip lengths, several other groups  have been performed MD simulations of the flow past striped anisotropic surfaces. Most of these studies considered a thick channel geometry, and focused only on calculating eigenvalues of the effective slip-length tensor~\cite{cottin.c:2004,Priezjev2005}. There are however two simulations which are directly relevant. Recent work~\cite{priezjev.n:2011} studied the dependence of a ratio of transverse
and forward effective slip velocities on the flow
orientation in a thick channel, but it did not explore the situation of a finite channel thickness. It has been found that the theoretical predictions based on a concept of tensorial slip
consistently overestimated the results of MD simulations, which has been attributed to the fact that the stripe width was
comparable to the molecular size. Other recent simulations~\cite{harting.j:2012} used the lattice-Boltzmann (LB) approach to calculate a flow in an \emph{asymmetric} channel (i.e. with one SH and one no-slip hydrophilic wall) of arbitrary thickness. This study confirmed that the eigenvalues of the effective slip-length tensor depend on the gap and validated the concept of tensorial slip. An excellent agreement of LB results with a continuous theory gave a strong support to macroscopic arguments. However, no attempts have been made to simulate the deflection of flow from the main direction, which for an asymmetric SH channel was not expected to be very dramatic~\cite{vinogradova.oi:2010}.

The goal of the present paper is to provide some general theoretical results to
guide the optimization of transverse hydrodynamic phenomena in a \emph{symmetric} channel with two aligned striped SH walls. In the case of \emph{thin} channels, symmetric striped walls provide
rigorous upper and lower bounds on the effective slip over all possible
two-phase patterns~\cite{feuillebois.f:2009} and are expected to generate very strong
transverse flow~\cite{mixer2010}, which could lead to very efficient mixing
upon spatial modulation of the texture. Here we consider a channel of
\emph{arbitrary} thickness, and optimize its orientation to maximize
pressure-driven transverse flow. Our consideration is based on
theoretical analysis and dissipative particle dynamics (DPD) simulations. By using relatively wide stripes as compared with a recent MD study~\cite{priezjev.n:2011} we find that our results in a very good quantitative
agreement with the macroscopic theory.

\section{Model and general consideration}

We consider a channel consisting of two equal symmetric SH walls located at
$y=0$ and $y=H$ and unbounded in the $x$ and $z$ directions as sketched in
Fig.\ \ref{fig:geometry}.
The origin of the coordinates $O$ is placed in the plane of a
liquid-gas interface at the center of the gas sector. The $x$ axis is defined
along the pressure gradient.  As in previous
publications~\cite{bocquet2007,belyaev.av:2010b,Priezjev2005,priezjev.n:2011,asmolov.es:2011},
we model SH plates as flat interfaces with no meniscus curvature, so that they
appear as being perfectly smooth with a pattern of boundary conditions. The latter
are taken as no-slip ($b_1=0$) over solid/liquid areas and as partial slip
($b_2=b$) over gas/liquid regions. In this idealization, by assuming a flat interface, we neglect an additional
mechanism for a dissipation connected with the meniscus
curvature~\cite{lauga2009,sbragaglia.m:2007}.
The fraction of the solid/liquid areas is denoted as $\phi_1$, and that of
the gas/liquid areas as $\phi_2 = 1 - \phi_1$.

The flow is governed by the Stokes equations
\begin{equation}\label{NS}
   \eta\nabla^2\textbf{v}=\nabla p,\,\,\,
  \nabla\cdot\textbf{v}=0,
\end{equation}
where $\textbf{u}$ is the velocity vector, and the average pressure gradient is
always aligned with the $x$-axis direction:
\begin{equation}\label{PressGrad}
   \langle\nabla p\rangle = (-\sigma, 0, 0)
\end{equation}
The local slip boundary conditions at the walls have the form
\begin{equation}\label{BC1}
   {\bf v}(x,0,z) = b(x,z)\cdot\frac{\partial {\bf v}}{\partial y}(x,0,z), \quad\hat{{\bf y}}\cdot {\bf v}(x,0,z) = 0.
\end{equation}
In addition, a symmetry condition applies at the mid-channel,
\begin{equation}\label{BC2}
   \frac{\partial\textbf{v}}{\partial y}(x,H/2,z)=0,\quad\hat{{\bf y}}\cdot {\bf v}(x,H/2,z) = 0.
\end{equation}
Here the local slip length $b(x,z)$ at the SH surface is generally
a function of both lateral coordinates.

The effective slip length $b_{\rm eff}$ at the SH
surface is defined as usual,
\begin{equation} \label{definition}
   b_{\rm eff}=\frac{\langle v_s\rangle}{\left\langle \left(\frac{\partial v}{\partial y}\right)_s \right\rangle},
\end{equation}
where $\langle \ldots\rangle$ denotes the average value in the plane $xOz$.
It can equivalently be defined via a permeability tensor
\begin{equation}\label{k_b}
   \textbf{k}= \frac{H^3}{12}\left(\textbf{I}+6 \frac{\textbf{b}_{\rm eff}}{H}\right)
\end{equation}

In the linear response regime, the averaged flow rate, $\langle\textbf{Q}\rangle$, is
proportional to $\langle \nabla p \rangle$ via the permeability tensor,
$\textbf{k}$:
\begin{equation}\label{Darcy_tensor}
    \langle\textbf{Q}\rangle = -\frac{1}{\eta} \textbf{k}\cdot \langle \nabla p \rangle,
\end{equation}
which may be rewritten as
\begin{eqnarray}\label{Q_II}
    \langle Q \rangle_x &=&\frac{\sigma}{\eta} \left(k^{\parallel}\cos^2\Theta+k^{\perp}\sin^2\Theta\right),\\
    \langle Q \rangle_z &=&\frac{\sigma}{\eta}\left(k^{\parallel}-k^{\perp}\right)\sin\Theta\cos\Theta.
\end{eqnarray}

Integrating the velocity profile across the channel we obtain
\begin{equation}\label{Q}
    \langle\textbf{Q}\rangle=\int\limits_0^H{\left\langle\textbf{v}(y)\right\rangle dy},
\end{equation}
or
\begin{eqnarray}
  \label{Qx}
  \langle Q \rangle_x &=& \int_0^H \langle v_x(y) \rangle dy, \\
  \label{Qz}
  \langle Q \rangle_z &=& \int_0^H \langle v_z(y) \rangle dy.
\end{eqnarray}

The above formula are valid regardless of the thickness of the symmetric SH
channel and are independent of the details of the textured surface. There could
be arbitrary patterns of local slip lengths, and the latter could itself be a
spatially varying tensor.

\section{Theory for striped patterns}

To illustrate the general model, in this section we focus on flat patterned
SH surfaces consisting of aligned periodic stripes, where the local (scalar) slip
length $b$ varies only in one direction. We mostly follow the approach developed
before~\cite{harting.j:2012},
but apply the method to a symmetric channel situation
as sketched in Fig.\ \ref{fig:geometry}, where the surfaces are covered with
arrays of gas/liquid stripes with width $\delta$ and period $L$.
Thus, $\phi_1 = 1-\delta/L$ and
$\phi_2=\delta/L$, respectively.

For transverse stripes, the flow is two-dimensional ${\bf v}=(v_x(x,y),
v_y(x,y),0)$, $v_x(x,0)=b(x) \partial_y v_x(x,0)$, and $v_y(x,0)=0$. For
longitudinal stripes, we also have a plane flow: ${\bf v}=(v_x(y,z), 0,0)$,
$v_x(0,z)=b(z)\partial_y v_x(0,z)$.  As the problem is linear in \textbf{v}, we
seek for a solution of the form
\begin{equation}
   \textbf{v}=\textbf{v}^{(0)}+\textbf{v}^{(1)},
\end{equation}
where $\textbf{v}^{(0)}$ is the velocity of the usual no-slip parabolic Poiseuille
flow
\begin{equation}
   \textbf{v}^{(0)}=\left(\frac{\sigma}{2\eta}y (H-y), 0, 0 \right),
\end{equation}
and $\textbf{v}^{(1)}$ is the SH slip-driven superimposed flow.

\subsection{Longitudinal stripes}
In this situation the problem is homogeneous in $x$-direction
($\partial/\partial x=0$). The slip length $b(x,z)=b(z)$ is periodic in $z$
with period $L$. The elementary cell is determined as $b(z)=b$ at
$|z|\leq\delta/2$, and $b(z)=0$ at $\delta/2<|z|\leq L/2$.  In this case the
velocity $\textbf{v}^{(1)}=(v_x^{(1)}, 0, 0)$ has only one nonzero component, which can
be determined by solving the Laplace equation
\begin{equation}\label{BP_u1_longit}
   \nabla^2 v_x^{(1)}(y,z)=0,
   \end{equation}
with the conditions (\ref{BC1})-(\ref{BC2}).

The Fourier expansion of a periodic solution satisfying (\ref{BC2}) reads
\begin{equation}\label{u1longit_2}
   v_x^{(1)}(y,z)=\frac{a_0}{2}+\sum^\infty_{n=1}a_n\cos(\lambda_n z) e^{-\lambda_n y} [1+ e^{\lambda_n (2y-H)}]
\end{equation}
with $\lambda_n=2\pi n/L$. The sine terms vanish due to symmetry.
Applying (\ref{BC1}) we then obtain a trigonometric dual series:
\begin{eqnarray}\label{DoubleSeries_a}
   \frac{\alpha_0}{2}+\sum^\infty_{n=1} \alpha_n \left[1+b \lambda_n \tanh(\lambda_n H/2) \right] \cos(\lambda_n z)=b ,\\ \nonumber |z|\leq \delta/2,   
\end{eqnarray}
\begin{equation}\label{DoubleSeries_b}
   \frac{\alpha_0}{2}+\sum^\infty_{n=1} \alpha_n \cos(\lambda_n z)=0,\quad \delta/2<|z|\leq L/2,
\end{equation}
where
\[
   \alpha_0=\frac{2\eta a_0}{\sigma H}; \quad \alpha_n=\frac{2\eta a_n(1+ e^{-\lambda_nH})}{\sigma H},\: n\geq 1.
\]
The dual series (\ref{DoubleSeries_a}), (\ref{DoubleSeries_b}) provide a
complete description of the hydrodynamic flow and the effective slip length
$b_{\rm eff}^{\parallel}=  \alpha_0 /2$ (due to Eq.\ (\ref{Q})) in the
longitudinal direction, given all the stated assumptions.  These equations can
be solved numerically, but exact results can be obtained in the limits of thin and
thick channels.

For a thin channel, $H\ll L$, we can use that $\left. \tanh t
\right|_{t\rightarrow0}=O(t)$. By substituting this expression into
(\ref{DoubleSeries_a}) and keeping only values of the first non-vanishing order, we find
\begin{equation}\label{beff_longit_smallH1}
   \left.b_{\rm eff}^{\parallel}\right|_{H \ll L} \simeq  \frac{1}{L} \int\limits_{-\delta/2}^{\delta/2} { b  dz} = \phi_2 b.
\end{equation}

This is an exact solution, representing a rigorous upper Wiener bound
on the effective slip over all possible two-phase patterns in a thin channel~\cite{feuillebois.f:2009}.

In the limit of a thick channel, $H\gg L$, we can use that
$\tanh(t\rightarrow \infty)\rightarrow 1$ and the dual series
(\ref{DoubleSeries_a})-(\ref{DoubleSeries_b}) can be solved exactly,
giving~\cite{belyaev.av:2010a}
\begin{equation}\label{beff_par_largeH}
  b_{\rm eff}^{\parallel} \simeq \frac{L}{\pi} \frac{\ln\left[\sec\left(\displaystyle\frac{\pi \phi_2}{2 }\right)\right]}{1+\displaystyle\frac{L}{\pi b}\ln\left[\sec\displaystyle\left(\frac{\pi \phi_2}{2 }\right)+\tan\displaystyle\left(\frac{\pi \phi_2}{2}\right)\right]}.
\end{equation}

\subsection{Transverse stripes} To describe a two-dimensional flow in this case
we use a standard technique and introduce a stream function $\psi(x,y)$ and the
vorticity vector $\boldsymbol{\omega} (x,y)$. Thus, the velocity field is
represented by $\textbf{v}(x,y)=\left(\partial \psi/\partial y, -\partial \psi/\partial x, 0 \right)$,
and the vorticity vector, $ \boldsymbol {\omega} (x,z)=\nabla\times\textbf{v}=(0, 0,\omega)$,
has only one nonzero component, which is equal to
\begin{equation}\label{vorticity}
   \omega=-\nabla^2\psi.
\end{equation}
The solution can then be presented as the sum of the base flow with homogeneous
no-slip condition and its perturbation caused by the presence of stripes as
\begin{eqnarray}\label{SF01}
   \psi &=& -\frac{\sigma}{\eta}\frac{y^3}{6}+\frac{\sigma H}{\eta}\frac{y^2}{4}+\psi_1, \\
   \omega &=& \frac{\sigma}{\eta}y-\frac{\sigma H}{2\eta}+\omega_1,
\end{eqnarray}

The problem for $\psi_1$ and $\omega_1$ reads
\begin{equation}\label{eq_psi1}
   \nabla^2\psi_1=-\omega_1,\quad \nabla^2\omega_1=0,
\end{equation}
which can be solved with respect to the BC (\ref{BC1})-(\ref{BC2})
and an additional condition that reflects our definition of the stream function:
\begin{equation}
\psi_1(x,0) = 0
\end{equation}
The general solution reads
\begin{eqnarray}\label{GenSol_psi1}
   &&\psi_1(x,y)=P_0 y\nonumber\\ &+&
   \sum^\infty_{n=1} \left(P^{(1)}_n - \frac{M^{(1)}_n}{2} \frac{y}{\lambda_n}\right)e^{\lambda_n y}\cos{\lambda_n x} \nonumber\\ &+&
   \sum^\infty_{n=1}\left(P^{(2)}_n + \frac{M^{(2)}_n}{2} \frac{y}{\lambda_n}\right)e^{-\lambda_n y}\cos{\lambda_n x},    
\end{eqnarray}
\begin{equation}\label{GenSol_omega1}
   \omega_1(x,y) = \sum^\infty_{n=1} \left(M^{(1)}_n e^{\lambda_n y} +M^{(2)}_n e^{-\lambda_n y}\right)\cos(\lambda_n x).
\end{equation}
The following relations between Fourier coefficients may be established:
\[
    P^{(1)}_n=-P^{(2)}_n\equiv -P_n,
\]
\[
    M^{(1)}_n=-\frac{4\lambda_n}{H}\sinh(\lambda_n H/2) e^{-\lambda_n H/2} P_n,
\]
\[
    M^{(2)}_n=\frac{4\lambda_n}{H}\sinh(\lambda_n H/2) e^{\lambda_n H/2} P_n.
\]
Again we obtain a dual series problem, which is similar to
(\ref{DoubleSeries_a}) and (\ref{DoubleSeries_b}):
\begin{eqnarray}\label{DoubleSeries_c}
      \frac{\alpha_0}{2}+\sum^\infty_{n=1} \alpha_n \left[1+2 b \lambda_n W(\lambda_n H) \right] \cos(\lambda_n x)=b,\\ \nonumber 0<x\leq \delta/2,   
\end{eqnarray}
\begin{equation}\label{DoubleSeries_d}
      \frac{\alpha_0}{2}+\sum^\infty_{n=1} \alpha_n \cos(\lambda_n x)=0,\quad \delta/2<x\leq L/2,
\end{equation}
Here, 
\begin{eqnarray}
   \frac{\alpha_0}{2} &=& b_{\rm eff}^{\perp}= \frac{2\eta}{\sigma H} P_0; \\
   \alpha_n &=& \frac{2\eta}{\sigma H} \cdot  2\left(\frac{\sinh(\lambda_n H)}{H}-\lambda_n\right) P_n, \: n\geq 1,
\end{eqnarray}
and
\begin{equation}\label{V}
   W(t)=\frac{\cosh(t)-1}{\sinh(t)-t}.
\end{equation}

In the limit of a thin channel
one has $W(t)|_{t\to \infty} \simeq 3t^{-1}+O(t)$,
from which it follows that
\begin{equation}\label{beff_longit_smallH}
  \left. b_{\rm eff}^{\perp}\right|_{H \ll L} = \frac{b H \phi_2}{H + 6b \phi_1}.
\end{equation}
This exact equation represents a rigorous lower Wiener bound on the effective slip over all possible two-phase patterns in a thin symmetric channel~\cite{feuillebois.f:2009}.

For completeness we discuss again the two limiting situations:
\begin{equation}\label{yy2}
   b_{\rm eff}^{\perp}|_{H\ll b, L} \simeq \frac{1}{6}\frac{\phi_2}{\phi_1}H  \:  \propto H,
\end{equation}
\begin{equation}\label{xx2}
   b_{\rm eff}^{\perp}|_{b\ll H\ll L} \simeq b\phi_2  \:  \propto b.
\end{equation}

In the thick channel limit, the dual series (\ref{DoubleSeries_c}) and
(\ref{DoubleSeries_d}) take the same form as in prior
work~\cite{vinogradova.oi:2010} (due to $W(x\rightarrow\infty)\rightarrow1$),
whence we derive~\cite{belyaev.av:2010a}
\begin{equation}\label{beff_ort_largeH}
  b_{\rm eff}^{\perp} \simeq \frac{L}{2 \pi} \frac{\ln\left[\sec\left(\displaystyle\frac{\pi \phi_2}{2 }\right)\right]}{1+\displaystyle\frac{L}{2 \pi b}\ln\left[\sec\displaystyle\left(\frac{\pi \phi_2}{2 }\right)+\tan\displaystyle\left(\frac{\pi \phi_2}{2}\right)\right]}.
\end{equation}

\subsection{Transverse flow}

The tensorial nature of the effective slip is physically
due to secondary flows transverse to the direction of the
applied pressure gradient. Now we focus on transverse
flow optimization, which is necessary for a passive mixing
in a symmetric SH channel. Our aim is to optimize the
texture, channel thickness and the angle $\Theta$ between the
directions of stripes and the pressure gradient, so that
$|\langle Q_z\rangle/\langle Q_x\rangle|$ is as large as possible.

Following the approach \cite{vinogradova.oi:2010}, one can derive
\begin{equation}
  \label{eq:q_ratio}
  F\equiv 
  \frac{ \langle Q \rangle_z } { \langle Q \rangle_x } = \frac{ 6(b_{\rm eff}^{\parallel}-b_{\rm eff}^{\perp}) \sin\Theta \cos\Theta} { H + 6 b_{\rm eff}^{\parallel} \cos^2 \Theta + 6 b_{\rm eff}^{\perp} \sin^2 \Theta},
\end{equation}
Consider now the tilt angle, $\Theta$, confined between  $0$ and $\pi/2$
and rewrite Eq.\ (\ref{eq:q_ratio}) as
\begin{equation}
  \label{eq:q_ratio_1}
  F \equiv
  \frac{ \langle Q \rangle_z } { \langle Q \rangle_x } = \frac{ 6 (\xi^2-1) \tau } { (h^2+6 \xi^2) +  (h^2+6)\tau^2 },
\end{equation}
where $\tau=\tan{\Theta}$, $\xi^2=b_{\rm eff}^{\parallel}/b_{\rm eff}^{\perp}$ and $h^2=H/b_{\rm eff}^{\perp}$. By
evaluating $\partial F/\partial \tau =0 $, we find that the maximum occurs at
\begin{equation}
 \label{t_star}
  \tau_* \equiv \tan{\Theta_*} = \left(\frac{ h^2 + 6 \xi^2  } { h^2 +6 } \right)^{1/2},
\end{equation}
and its value is
\begin{equation}
 \label{F_star}
  F(\tau_*) = \frac{ 3 (\xi^2 -1) } { [ (h^2+6 \xi^2) (h^2 +6) ]^{1/2} }.
\end{equation}

In the limit of a \emph{thick} channel, $H\gg L$, $h \rightarrow \infty$ (owing to the fact that $b_{\rm eff}^{\perp}$ is limited by the local slip length $b$ and independent of $H$ for that case), and Eq.\ (\ref{t_star}) gives
\begin{equation}
 \label{t_star_thick}
 \left. \tan{\Theta_*} \right|_{H \gg L} \simeq  1, \quad  \left. \Theta_*\right|_{H\gg L} \simeq \pi/4,
\end{equation}
and, correspondingly,
\begin{equation}
 \label{F_star_thick}
 \left. F(\tau_*) \right|_{H/L\rightarrow \infty} \simeq \frac{3(\xi^2-1)}{h^2} = O(L/H).
\end{equation}
Therefore, the mixing in a thick symmetric SH channel
would be not very efficient. According to our formula,
to maximize $|\langle Q_z\rangle/\langle Q_x\rangle|$ an efficient strategy would be to
use striped textures with largest physically possible $b$ and
a very low fraction of a solid phase, $\phi_2\rightarrow 1$.

When the channel is thin, $H\ll L$, the interplay between $b$ and $H$ gives two possibilities.
First, if $b\ll H\ll L$, the slip length is isotropic,
$b_{\rm eff}^{\perp}=b\phi_2 (1-6\phi_1 b/H) \simeq b_{\rm eff}^{\parallel}$, and $h^2=H/(b\phi_2) \gg 1$, $\xi \simeq 1$. Therefore, again $\Theta_* \simeq \pi/4$, yet the value
\begin{equation}
  F(\tau_*) \simeq \frac{18 \phi_1\phi_2 b^2}{H^2} = O\left(\left[\frac{b}{H}\right]^2\right)
\end{equation}
is negligible.
Then, for $H\ll \min\{b, L\}$ we get
\[
   h^2 \simeq 6\phi_2/\phi_1= {\rm const},
\]
\[
   \xi^2 = 6 b \phi_1 /H  \gg 1.
\]
Substitution of these expressions into Eqs.\ (\ref{t_star}) and  (\ref{F_star}) gives
\begin{equation} \label{t_star_thin2}
   \left. \tan{\Theta_*} \right|_{H \ll \min\{b,L\}} \simeq  \left(\frac{6 b\phi_1\phi_2}{H}\right)^{1/2},
\end{equation}
\begin{equation} \label{F_star_thin2}
   \left. F(\tau_*) \right|_{H \ll \min\{b,L\}} \simeq  \frac{1}{2}\left(\frac{6 b\phi_1\phi_2}{H}\right)^{1/2} \gg 1.
\end{equation}
Since $\phi_1\phi_2=(1-\phi_2)\phi_2$ is maximal for $\phi_2=\phi_1=0.5$,
the direction of optimal inflow angle is
\begin{equation} \label{t_star_thin3}
    \left. \Theta_*\right|_{H \ll \min\{b,L\}} \simeq \frac{\pi}{2} - \left(\frac{2 H}{3 b}\right)^{1/2},
\end{equation}
which coincides with the results of~\cite{mixer2010} derived by using a different method.

\section{Simulation method}
\label{sec:simulation}

Our simulations are done using Dissipative Particle Dynamics (DPD), an
established method for mesoscale fluid simulations, which is fully off-lattice
and particle based and naturally includes thermal fluctuations
\cite{Koelman1993, Espanol1995, Groot1997}. More specifically, we use a DPD
version without conservative interactions, and combine that with a tunable-slip
method that allows one to implement arbitrary hydrodynamic boundary conditions
\cite{Smiatek2008}.  The detailed implementation of the tunable-slip method can
be found in Ref.~\cite{Smiatek2008}. In the following we only give a brief
description and introduce the simulation parameters.

In the tunable-slip boundary approach, the interaction between the channel
walls and the fluid particles has two contributions. The first is a
Weeks-Chandler-Andersen (WCA) interaction to mimic the impermeable surfaces,
\begin{equation}
u (y) =  \left\{ \begin{array}{cl} 4 \epsilon [ (\frac{\sigma}{y})^{12}
  - (\frac{\sigma}{y})^6 + \frac{1}{4} ],  \quad & y < 2^{1/6} \sigma \\
0, & y \ge 2^{1/6} \sigma \end{array} \right.
\end{equation}
where $y$ is the distance between the fluid particle and the wall.  This is
just a Lennard-Jones interaction with a cutoff at the potential minimum,
corresponding to the pure repulsive part of the potential. In the following,
the WCA parameters will set our simulation units, {\em i.e.}, the energy unit
$\epsilon$ and the length unit $\sigma$.  The third unit is the particle mass
$m$. The second part of the wall-fluid interaction is a coarse-grained friction
force, which is introduced in a similar spirit than the DPD approach.
Specifically, the effect of the wall friction on $i$-th particle is implemented
by introducing a pair of Langevin-type forces,
\begin{equation}
  \mathbf{F}_i^{wall} = \mathbf{F}_i^D + \mathbf{F}_i^R .
\end{equation}
The dissipative contribution has the form
\begin{equation}
  \mathbf{F}_i^D = - \gamma_L \omega_L(z) (\mathbf{v}_i - \mathbf{v}_{wall}).
\end{equation}
This force is proportional to the relative fluid velocity with respect of the
wall, and the proportionality factor is the local viscosity $\gamma_L \omega_L(z)$.
The position-dependent function $\omega_L(z)$ is a monotonically decreasing
function of the wall-particle separation, and vanishes when the fluid particle
is further away from the wall than a cutoff distance $z_c$. In our simulations,
we take $\omega_L(z)$ to decrease linearly from 1 to 0. The prefactor $\gamma_L$
characterizes the strength of the wall friction and can be used to tune the
value of the slip length.  A random force obeying the fluctuation-dissipation
theorem is required to ensure the correct equilibrium statistics,
\begin{equation}
  \mathbf{F}_i^R =  \boldsymbol{\xi}_i \sqrt{2 k_B T \gamma_L \omega_L(z)},
\end{equation}
where each component of $\boldsymbol{\xi}_i$ is a Gaussian distributed random
variable with zero mean and unit variance.

The simulations are carried out using the open source simulation package
ESPResSo \cite{Limbach2006}.  All simulations are performed with a time step
$\Delta t = 0.01 \sqrt{m/\epsilon}\sigma$, and the temperature of the system is
set at $k_BT=1\epsilon$.  The fluid has a density $\rho = 3.75 \sigma^{-3}$.
Fluid particles have no conservative interactions, they interact only with the
dissipative part of the DPD interactions, The DPD interaction parameter is
chosen at $\gamma_{DPD} = 5.0 \sqrt{m\epsilon}/\sigma$ and the cutoff radius is
$1.0\sigma$, which results in a shear viscosity of $\eta_s = 1.35\pm0.01
\sqrt{m\epsilon}/\sigma^2$.

\begin{figure}[htbp]
  \includegraphics[width=1.0\columnwidth]{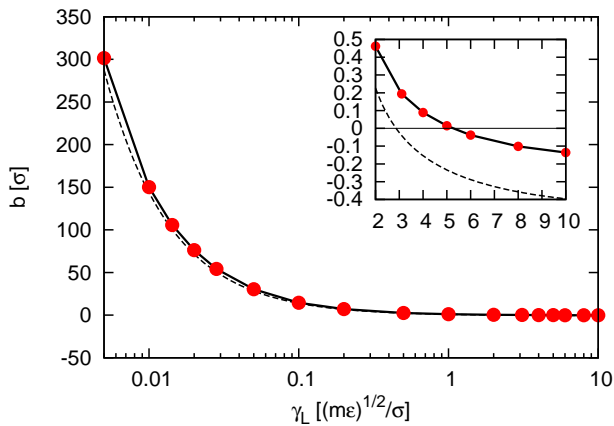}
  \caption{The relation between the slip length $b$ and the wall friction
   parameter $\gamma_L$, for a wall interaction cutoff $2.0\sigma$. The
   inset shows an enlarged portion of the region where the slip length
   is zero. The no slip boundary condition can be implemented by using
   $\gamma_L=5.26 \sqrt{m\epsilon}/\sigma$. Dashed curves show analytical predictions~\cite{Smiatek2008}.}
  \label{fig:b_gammaL}
\end{figure}

The slip length $b$ can be calculated analytically as a function of the
simulation parameters (the wall friction parameter $\gamma_L$ and cutoff $r_c$)
to a very good approximation.  For the purpose of this work,
however, we need highly accurate values for both $b$ and the position of the
hydrodynamic boundary, therefore we determined it
with the method described in Ref.~\cite{Smiatek2008} by simulating plane
Poiseuille flow and plane Couette flow.  Fig.\
\ref{fig:b_gammaL} shows the relation between the slip length and the wall
friction parameter $\gamma_L$, for a wall interaction cutoff $r_c=2.0\sigma$.
By setting the wall friction parameter $\gamma_L$, we can adjust the slip
length to arbitrary values. The no-slip boundary condition is implemented with
$\gamma_L=5.26 \sqrt{m\epsilon}/\sigma$.  The
position of the hydrodynamic boundary, was determined to be
$1.06\pm0.12\sigma$ away from the simulation wall.

For simulations of patterned surfaces, we use a box of
$20\sigma \times (H+2\sigma) \times 50\sigma$.
The two impermeable surfaces lie parallel to the $xz$-plane and are separated
by a distance $H+2\sigma$.
Periodic boundary conditions are assumed in the
$xz$ plane in the directions parallel and perpendicular to the stripes.  We
have tested that increasing the simulation box does not change the outcome.
The surfaces are decorated by alternating no-slip and partial slip stripes.
The pattern has a periodicity of $L=50\sigma$.  The wall friction parameters
are chosen $\gamma_L=5.26 \sqrt{m\epsilon}/\sigma$, which implements a no-slip
boundary condition, and $\gamma_L=0.02820 \sqrt{m\epsilon}/\sigma$,
corresponding to $b=50\sigma$.  A simulation system is completely specified by
the following parameters: ($L$, $b$, $\phi_2$, $H$).

The simulation starts with randomly distributed DPD particles.  An external
body force is assigned to each particle to mimic the pressure gradient. The
amplitude of the force is adjusted such that the maximum shear rate at the
surface is $\dot{\gamma} = 0.01 \sqrt{m/\epsilon \sigma^2}$. Small shear
rates are necessary to avoid shear-rate dependencies of the slip length
\cite{thompson.pa:1997} and reduce inertia effects. With these parameters,
typical Reynolds numbers in our system are still of order 10, which is much
larger than typical values in microfluidic setups. To reach realistic Reynolds
numbers, one would have to reduce the body force by four orders of magnitude.
Unfortunately, the necessary simulation time for gathering data with
sufficiently good statistics would then increase prohibitively, since the values of the
relevant observables become very small and the correlation times increase with
$1/\dot{\gamma}$. Test runs with selective parameters were performed using a smaller body force,
but the results did not change significantly.
Therefore, we chose to work at these relatively high
Reynolds numbers (we are still deeply in the laminar regime), and tolerate
slight inertia effects, which we will discuss further below.

\begin{figure}[tbp]
  (a)\includegraphics[width=0.9\columnwidth]{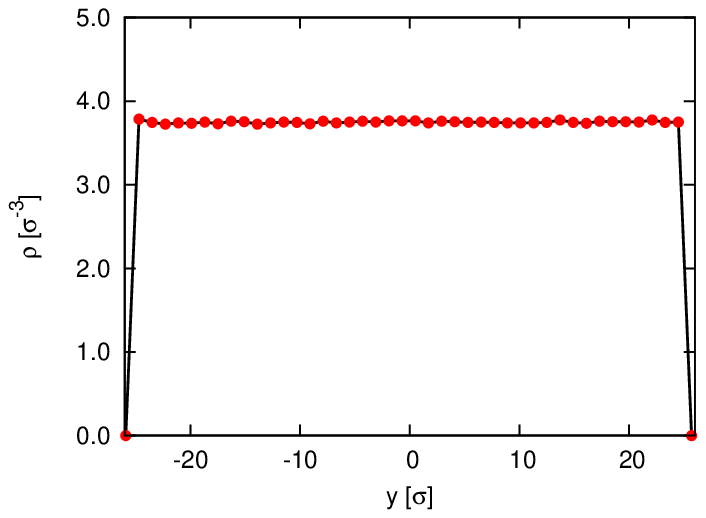}
  (b)\includegraphics[width=0.9\columnwidth]{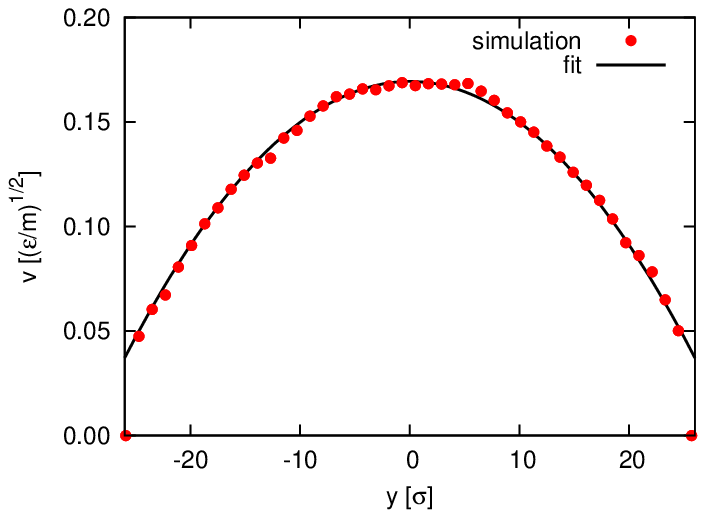}
  \caption{Typical density (a) and velocity  (b) profiles simulated for a
longitudinal flow and a texture with  $L=H=b=50\sigma$, $\phi_2=0.5$. }
  \label{fig:stripes}
\end{figure}

One important criterion for a correct simulation of the pressure-driven flow is
a constant density profile.  Fig.\ \ref{fig:stripes}(a) shows an example of the
fluid density profile for $\rho=3.75\sigma^{-3}$.  The channel has a thickness
of $H=50\sigma$.  The periodicity of the striped pattern is $L=50\sigma$, and
the partial-slip part has a slip length of $b=50\sigma$ and area fraction
$\phi_2=0.5$.  The fluid density is uniform in the middle of the channel.
and drops to zero in the
proximity of the wall due to the repulsive interaction of the wall on the fluid
particles.  This leads to a physical wall position which is approximately
$1.0\sigma$ away from the simulation wall.  The quoted value of channel
thickness $H$ is the separation between the physical walls. In contrast
to earlier MD simulations \cite{Priezjev2005}, there are no molecular
layering effects and density oscillations near the walls, because our
DPD particles have no conservative interactions.

Fig.\ \ref{fig:stripes}(b) shows the averaged velocity profile for the same
system. The system is allowed to run up to $5\times10^5$ time steps to reach a steady state, and the velocity profile is then averaged over long time intervals (up to $10^4$ time
steps) with horizontal bins of thickness $\Delta y = 0.02\sigma$.  Close to the
wall, the velocity is zero because no particles are present in that
region.  In the middle of the channel, the velocity profile exhibits a
parabolic shape, such that we can fit the velocity to a plane Poiseuille
flow and obtain an effective slip length.  The fitting function we used is
\begin{equation}
  v(y) = \frac{\rho F^{\rm ext}}{2\eta_s} ( y_B^2 - y^2 + 2b y_B),
\label{eq:poiseuille}
\end{equation}
where $F^{\rm ext}$ is the external force applied to each DPD particle to
induce the pressure gradient.  For the position of the hydrodynamic boundary
$y_B$, we use the value obtained from simulations for homogeneous surfaces
($1.06\pm0.12 \sigma$ from the simulation wall).  We assume that the position
of the hydrodynamic boundary remains the same for striped surfaces.
The values of $\langle Q \rangle_x$ and $\langle Q \rangle_z$ are then
computed with Eqs.\ (\ref{Qx})-(\ref{Qz}).

\section{Results and Discussion}
\label{sec:results}

In this section, we present the DPD simulation results and compare them with
predictions from the continuous theory.  Throughout the section, we use striped
patterns with periodicity $L=50 \sigma$ and slip length $b=50 \sigma$ on the
slippery areas. We study the effect of varying the film thickness $H$,
the angle $\Theta$ between the applied force and the stripes, and the
area fraction $\phi_2$ of slippery areas.

\begin{figure}[bthp]
  (a)\includegraphics[width=0.8\columnwidth]{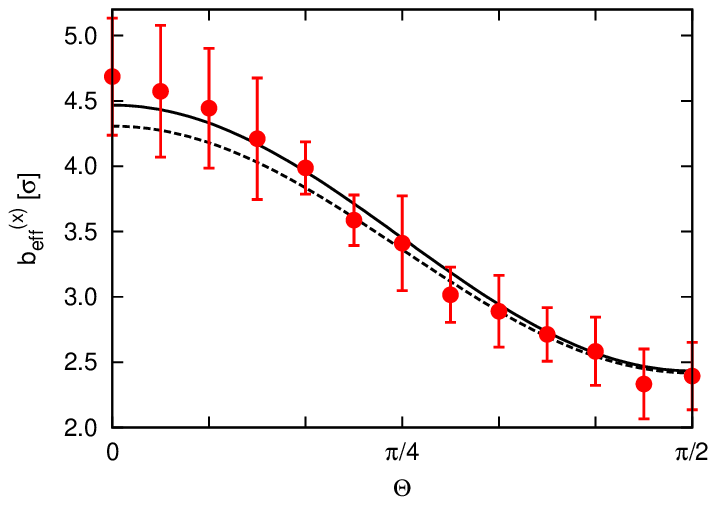}
  (b)\includegraphics[width=0.8\columnwidth]{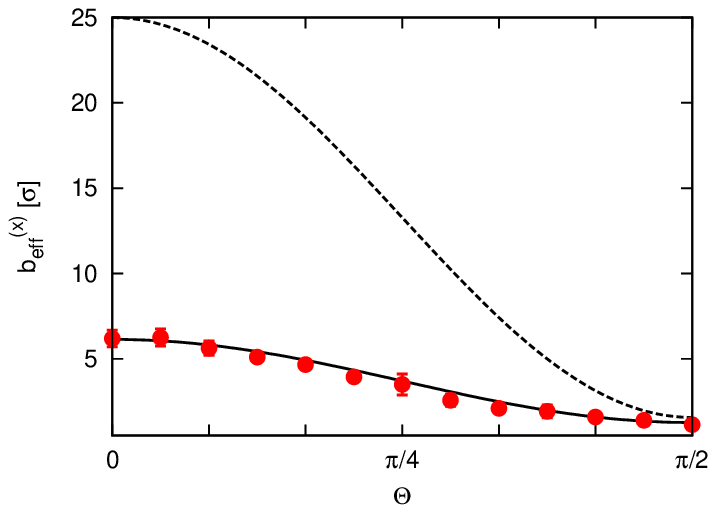}
  \caption{The effective downstream slip length, $b_{\rm eff}^{(x)}$, as a
function of tilt angle $\Theta$ for a pattern with $L=b=50\sigma$ and
$\phi_2=0.5$. Symbols are simulation data. Solid  lines are theoretical values
calculated using Eq.\ (\ref{eq:beff_xx}) with eigenvalues obtained by a
numerical solution of Eqs.\ (\ref{DoubleSeries_a}), (\ref{DoubleSeries_b}) and
(\ref{DoubleSeries_c}), (\ref{DoubleSeries_d}): (a) $H=50\sigma$. The thick
channel limit (dashed line) is calculated with Eqs.\ (\ref{beff_par_largeH}) and
(\ref{beff_ort_largeH}). (b) $H=10\sigma$. The thin channel limit (dashed
line), is calculated with Eqs.\ (\ref{beff_longit_smallH1}) and
(\ref{beff_longit_smallH})}
  \label{fig:theta_beff}

\end{figure}

We start with varying $\Theta$ in a system where the stick and slip areas are
equal, $\phi_2=0.5$, which corresponds to maximum transverse flow in a thin
channel situation \cite{mixer2010}.  Fig.\ \ref{fig:theta_beff} shows the
results for the effective downstream slip lengths, $b_{\rm eff}^{(x)}$, as
obtained from Poiseuille fits (\ref{eq:poiseuille}) to the $x$ component of the
velocity $v_x$, for two values of the channel thickness, $H=50 \sigma$ and
$H=10 \sigma$.  We emphasize that in both cases we formally have an
intermediate channel situation, since $H/L = O(1)$.  The error bars have been
obtained from averaging over five independent runs.  Since we model an
infinitely extended slit by virtue of applying periodic boundary conditions in
the $xz$ plane, the downstream slip length in our system, $b_{\rm eff}^{(x)}$,
corresponds to $b^{\rm eff}_{xx}$.  (In channels that are confined in the $y$
direction, a transverse pressure builds up that renormalizes $b_{\rm
eff}^{(x)}$\cite{bazant08}.) Fig.\ \ref{fig:theta_beff} also includes a
theoretical curve calculated using Eq.\ (\ref{beff_def1}), which can be
explicitly written as
\begin{equation}
  \label{eq:beff_xx}
  b^{\rm eff}_{xx} = b_{\rm eff}^{\parallel} \cos^2 \Theta
  + b_{\rm eff}^{\perp} \sin^2 \Theta .
\end{equation}
Here, the eigenvalues of the slip-length tensor are obtained by numerical
solution of the dual series, using the procedure described in Ref.\
\cite{harting.j:2012}.  Also included in Fig.\ \ref{fig:theta_beff} (dashed
curves) are the $xx$ components of the effective slip length tensor calculated
in the limit of thick [Fig.\ \ref{fig:theta_beff}(a)] and thin
[Fig.\ \ref{fig:theta_beff}(b)] channels.  These indicate the range of $b^{\rm
eff}_{xx}$ in this (symmetric) channel geometry.

The simulation data are in a good agreement with theoretical predictions,
confirming the anisotropy of the flow and the validity of the concept of a
tensorial slip for arbitrary channel thickness~\cite{harting.j:2012}.  Most
notably, the effective slip is larger for thinner channels, somewhat in
contrast to previous findings for asymmetric channels, where the magnitude of
the effective slip length in a thin gap was much smaller than that in a thick
channels ~\cite{harting.j:2012,asmolov.es:2011,belyaev.av:2010b}. This means
that transverse hydrodynamic phenomena should be enhanced in thinner channels.

\begin{figure}[bthp]
  (a)\includegraphics[width=0.9\columnwidth]{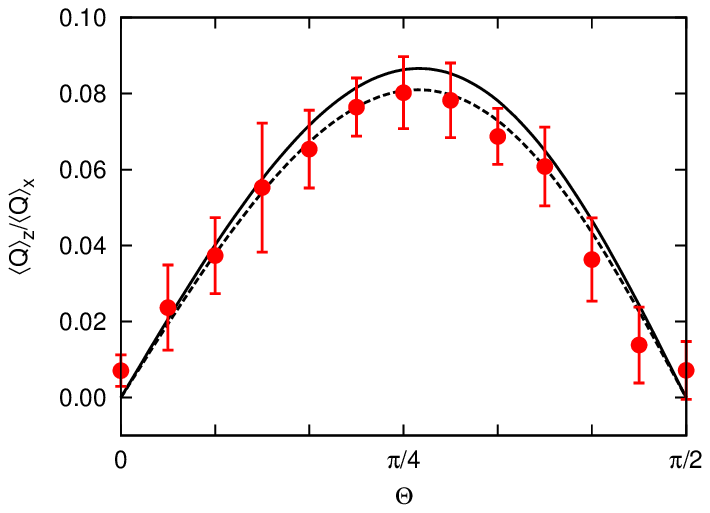}
  (b)\includegraphics[width=0.9\columnwidth]{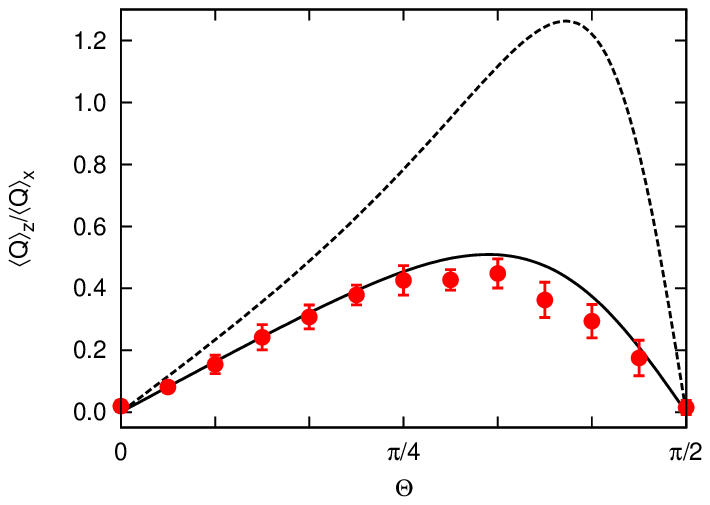}
  \caption{The ratio $\langle Q \rangle_z / \langle Q \rangle_x $ as a function
of the tilt angle $\Theta$ obtained with Eq.\ (\ref{eq:q_ratio}) for the data
sets from Fig.\ \ref{fig:theta_beff} for (a) $H=50 \sigma$, (b) $H=10 \sigma$.
Symbols are simulation data,  solid curves represent theoretical values, and
dashed curves show asymptotic predictions in the limit of thick channels (a)
and thin channels (b). }
  \label{fig:theta_q}
\end{figure}

The development of transverse flow is illustrated in Fig.\ \ref{fig:theta_q},
where the data for the ratio $\langle Q \rangle_z / \langle Q \rangle_x$ of the
measured averaged longitudinal and transverse flow rates at different $\Theta$
are presented and compared with the theoretical prediction of Eq.\
(\ref{eq:q_ratio}).  The simulation data and the theoretical prediction agree
quite well. There are some discrepancies at intermediate angles $\Theta$, where
the simulation data for $\langle Q \rangle_z / \langle Q \rangle_x$ are smaller
than predicted by theory. These slight deviations are most likely a result of
the finite Reynolds numbers, which are of order ${\cal O}(10)$ in the present
simulations as discussed in Section \ref{sec:simulation}.  When increasing the
bulk force and hence the average flow velocities, the deviations increase, thus
they will presumably vanish in the Stokes limit.  A pressure gradient in
eigendirections cannot produce any transverse flow (and the tensorial boundary
condition reduces to a scalar one), which is well seen in Fig.\
\ref{fig:theta_q}. In all other situations the direction of flow is different
from that of the pressure gradient. The maximum value of $\langle Q \rangle_z /
\langle Q \rangle_x$ is smaller for the larger channel (Fig.\
\ref{fig:theta_q}(a)).  In this case, this maximum is reached at
$\Theta=\pi/4$, which agrees with theoretical calculations made with Eq.\
(\ref{t_star_thick}) for a thick channel. For the thinner channel, however, the
maximal $\langle Q \rangle_z / \langle Q \rangle_x$ is observed at larger
$\Theta$ (Fig.\ \ref{fig:theta_q}(b)), also in agreement with the theory (see
discussion above).  An important conclusion from these results is that the surface
textures which optimize transverse flow differ significantly from those
optimizing effective (forward) slip. Similar predictions have been made
for other channel configurations ~\cite{vinogradova.oi:2010,mixer2010}.

\begin{figure}[tbhp]
  (a)\includegraphics[width=0.9\columnwidth]{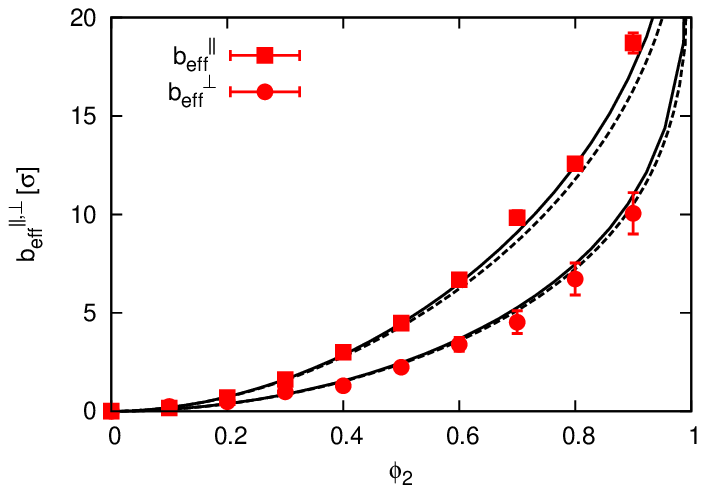}
  (b)\includegraphics[width=0.9\columnwidth]{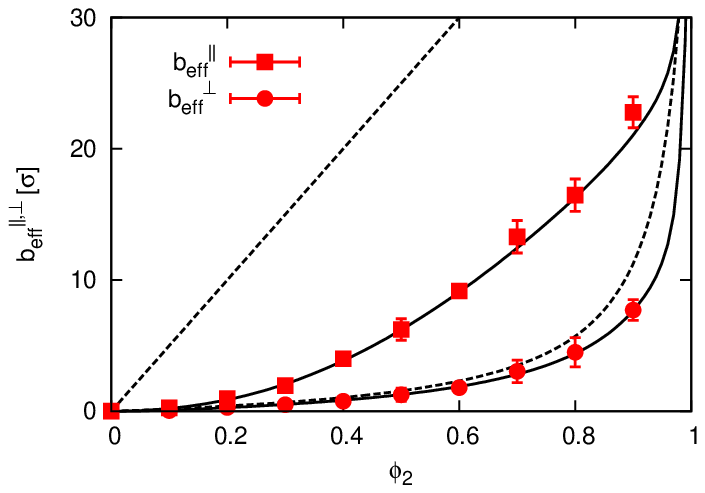}
 \caption{The eigenvalues of the effective slip length tensor (symbols) as
a function of $\phi_2$. Solid curves are theoretical values obtained by a
numerical solution of Eqs.\ (\ref{DoubleSeries_a}), (\ref{DoubleSeries_b}) and
(\ref{DoubleSeries_c}), (\ref{DoubleSeries_d}). Calculations were made for a
pattern with $L=b=50\sigma$:  (a) $H=50\sigma$. Dashed curves are calculated
with Eqs.\ (\ref{beff_par_largeH}) and (\ref{beff_ort_largeH})  (b)
$H=10\sigma$. Dashed curves are computed with Eqs.\ (\ref{beff_longit_smallH1})
and (\ref{beff_longit_smallH}).}
  \label{fig:phi2_beff}

\end{figure}

Next we examine the effect of varying the fraction of slippery gas/liquid
phase, $\phi_2$.  Fig.\ \ref{fig:phi2_beff} shows the eigenvalues, $b_{\rm
eff}^{\parallel}$ and $b_{\rm eff}^{\perp}$, of the slip-length tensor as a
function of $\phi_2$.  The two data sets again correspond to different
thickness of the channel, $H=50\sigma$ and $H=10\sigma$. The results clearly
demonstrate that  $\phi_2$ is the main factor determining the value of
effective slip, which significantly increases with the fraction of the slippery
sectors.  The theoretical curves match the simulation data very nicely.

\begin{figure}[tbh]
  (a)\includegraphics[width=0.9\columnwidth]{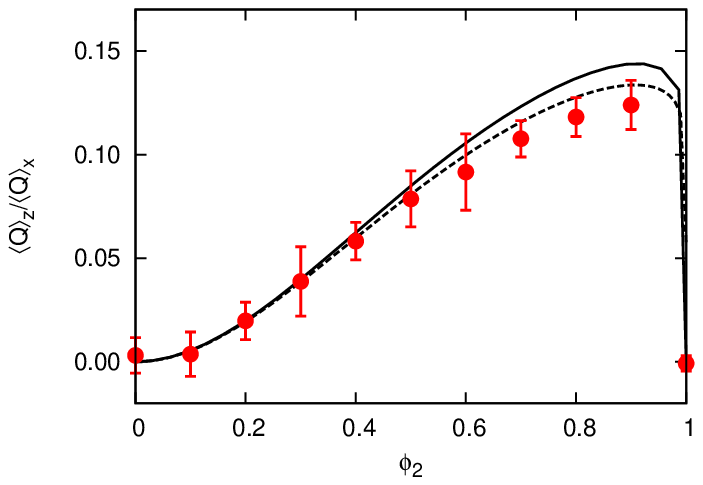}\\
  (b)\includegraphics[width=0.9\columnwidth]{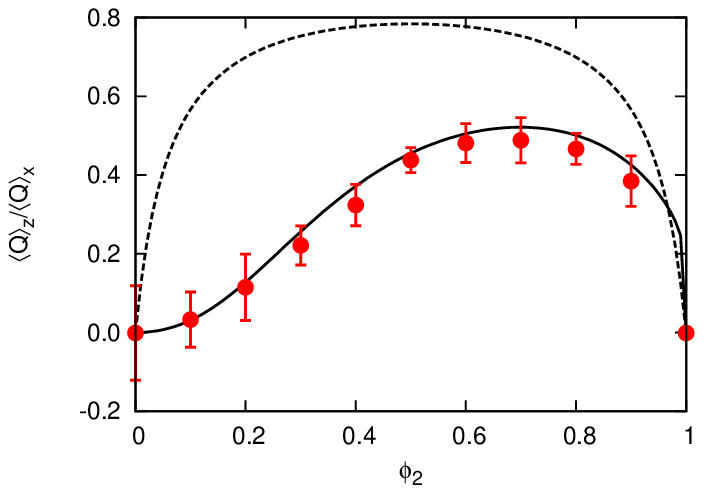}\\
  \caption{The ratio $\langle Q \rangle_z / \langle Q \rangle_x $ as a function
of $\phi_2$ obtained with Eq.\ (\ref{eq:q_ratio}) for $\Theta=\pi/4$ by using
the data sets from Fig.\ \ref{fig:phi2_beff} (a) $H=50 \sigma$, (b) $H=10 \sigma$.
Symbols are simulation data, solid curves represent theoretical values, and
dashed curves show asymptotic predictions in the limit of thick channels (a)
and thin channels (b). }
  \label{fig:phi2_q}
\end{figure}

 To illustrate the effect of $\phi_2$ on the transverse phenomena, we now fix
$\Theta=\pi/4$ and measure $\langle Q \rangle_z / \langle Q \rangle_x$. The
simulation data presented in Fig.\ \ref{fig:phi2_q} show that the maximum value
of $\langle Q \rangle_z / \langle Q \rangle_x$ is observed at  $\phi_2 \ge 0.5$
and depends on the channel thickness. As discussed above,  $\phi_2 = 0.5$
results in maximum transverse flow in a thin channel (thin compared to the
period of the texture)~\cite{mixer2010}. For thick channels, the continuum
theory predicts maximal transverse flow at $\phi_2\rightarrow 1$. The
simulation data, obtained for two values of $H$, confirms these trends. Fig.\
\ref{fig:phi2_q} also illustrates that transverse phenomena are more efficient
in a thinner channel. The data sets in Fig.\ \ref{fig:phi2_q}  are in
quantitative agreement with predictions of the continuum theory. Only for the
thick channels and $\phi_2 \ge 0.5$ do we observe slight deviations, which
again presumably reflect inertia effects.

\begin{figure}[tbhp]
  (a)\includegraphics[width=0.9\columnwidth]{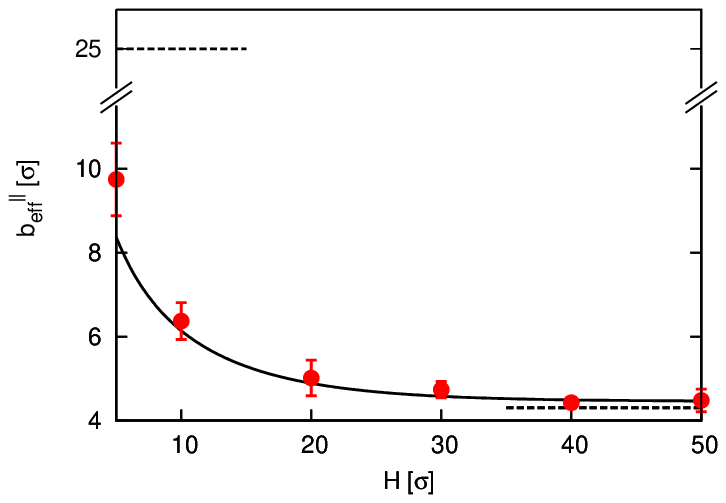}
  (b)\includegraphics[width=0.9\columnwidth]{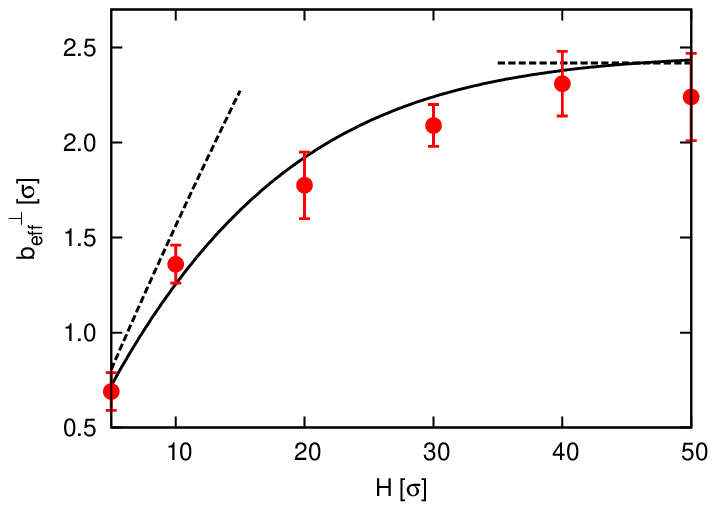}
  \caption{The longitudinal (a) and transverse (b) effective slip lengths as a
function of the channel height $H$ for a texture with $L=b=50\sigma$ and
$\phi_2=0.5$. The theoretical curves are obtained by a numerical solution of
Eqs.\ (\ref{DoubleSeries_a}), (\ref{DoubleSeries_b}) and
(\ref{DoubleSeries_c}), (\ref{DoubleSeries_d}). Dashed lines show expected
asymptotics in the limit of thin and thick channels.}
  \label{fig:H_beff}
\end{figure}

The simulation and theoretical data presented in Figs.\ \ref{fig:theta_beff}
and \ref{fig:phi2_beff} show that the eigenvalues of the slip-length tensor
depend on the channel thickness $H$, in agreement with earlier
predictions~\cite{harting.j:2012}. To examine this dependence in more detail,
we now fix $\phi_2=0.5$ and vary $H$ in a large range.  The smallest gap
was taken to be  $H=5\sigma$, for smaller gaps the velocity profile can no
longer be resolved satisfactorily. The largest thickness was chosen  to be
equal to $L=50\sigma$. The eigenvalues of the slip-length tensor are shown in
Fig.\ \ref{fig:H_beff}.  Also included are continuum theoretical calculations
(solid curves) and asymptotic values (dashed curves) expected in the true
limits of thin and thick channels. The data show that the longitudinal
effective slip decreases  with $H$.  In contrast, the transverse effective slip
increases with $H$, so that the difference between two eigenvalues is largest
for thin channels.  We remark and stress that the effective slip reaches the
asymptotic values predicted for a thick channel already at $H=O(L)$. A similar
observation was made in~\cite{harting.j:2012} for asymmetric channels. This
result is remarkable since it suggests that the thick channel limit, where the
effective slip is a property of single interfaces and does not depend on $H$,
is already reached for channels whose thickness is of the order of the texture
period. In practice, this implies that huge simulations with large simulation
boxes are not necessary to determine the effective hydrodynamic behavior in
such systems.  We note however that even for the smallest gaps considered here
($H/L=0.1$), our theoretical and simulation results still deviate strongly from
the upper Wiener bound, Eq.\ (\ref{beff_longit_smallH1}), although they are
very close to the lower Wiener bound, Eq.\ (\ref{beff_longit_smallH}).

\begin{figure}[htbp]
  \includegraphics[width=0.9\columnwidth]{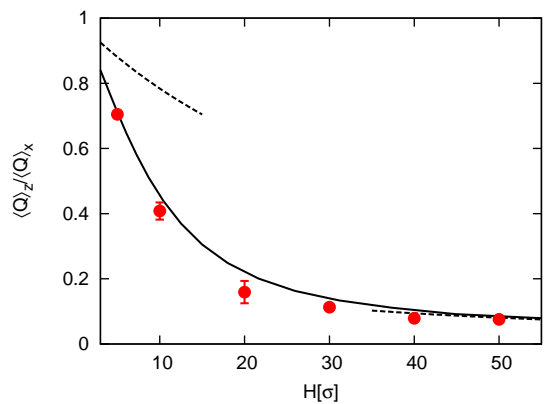}
    \caption{The ratio between the transverse and longitudinal flow rates
$\langle Q \rangle_z / \langle Q \rangle_x$  as a function of channel thickness
$H$ for a pattern with $L=b=50\sigma$, $\phi_2=0.5$, and $\Theta=\pi/4$.
Symbols are simulation data and the lines represent theoretical values obtained
using Eq.\ (\ref{eq:q_ratio}). Dashed lines show expected asymptotics in the
limit of thin and thick channels.}
  \label{fig:H_q}
\end{figure}

Fig.\ \ref{fig:H_q} shows the corresponding simulation and theoretical data for
$\langle Q \rangle_z / \langle Q \rangle_x$ as a function of channel thickness
$H$.  The stripe texture is the same as in Fig.\ \ref{fig:H_beff}, i.e.,
$\phi_2=0.5$, and the tilt angle was chosen $\Theta=\pi/4$. As discussed above,
this angle leads to a maximal transverse flow in case of a thick channel, but
not in the thin channel situation, where a much larger angle is required to
optimize the transverse phenomena. Nevertheless, $\langle Q \rangle_z / \langle
Q \rangle_x$ increases dramatically, when the channel becomes thinner.  With
our range of parameters, however, we are still well below the limiting value of
$\langle Q \rangle_z / \langle Q \rangle_x$, which would be expected in case
Wiener bounds were attained.

\begin{figure*}[t!]
  (a)\includegraphics[width=0.45\columnwidth]{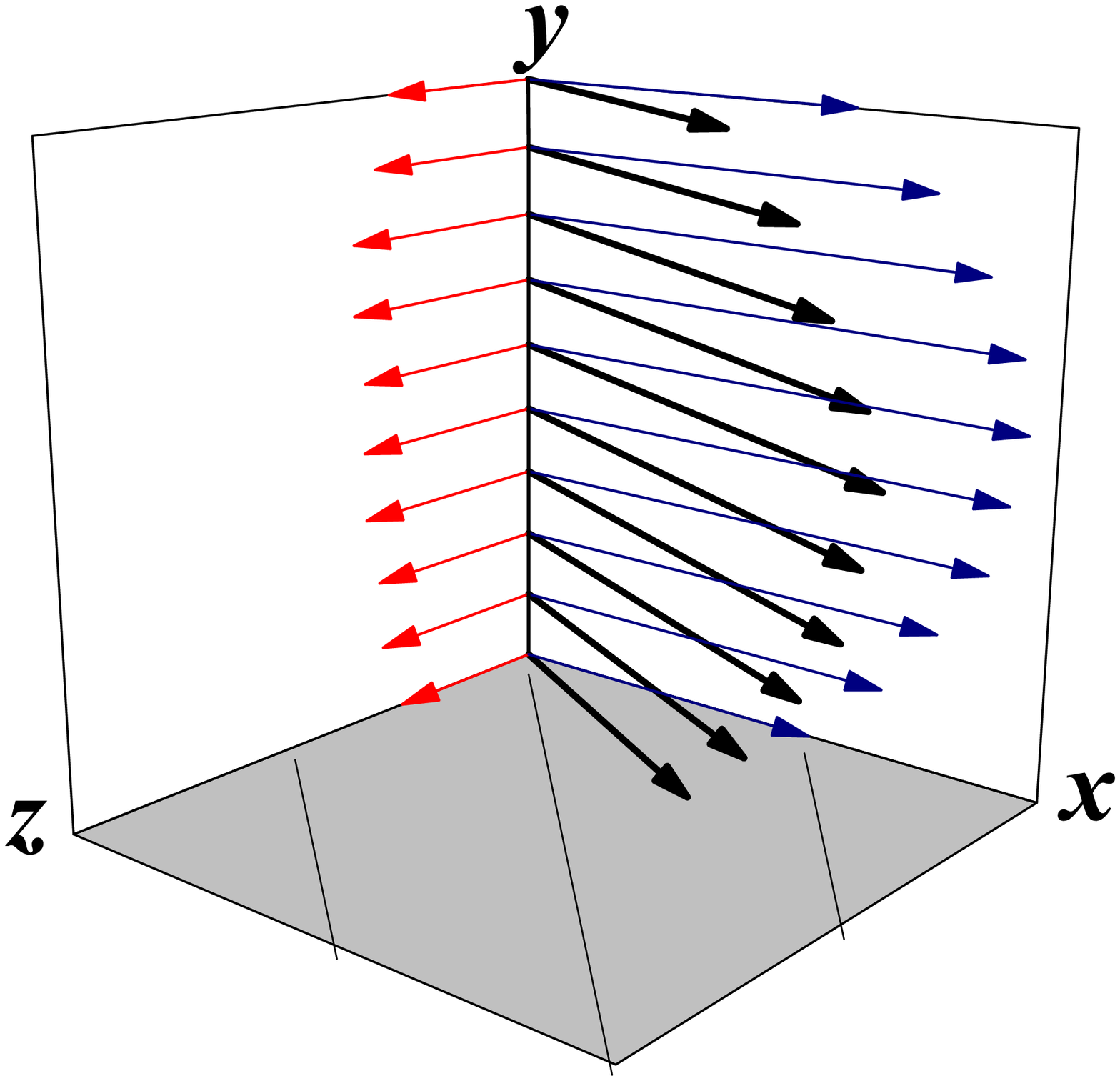}
  (b)\includegraphics[width=0.45\columnwidth]{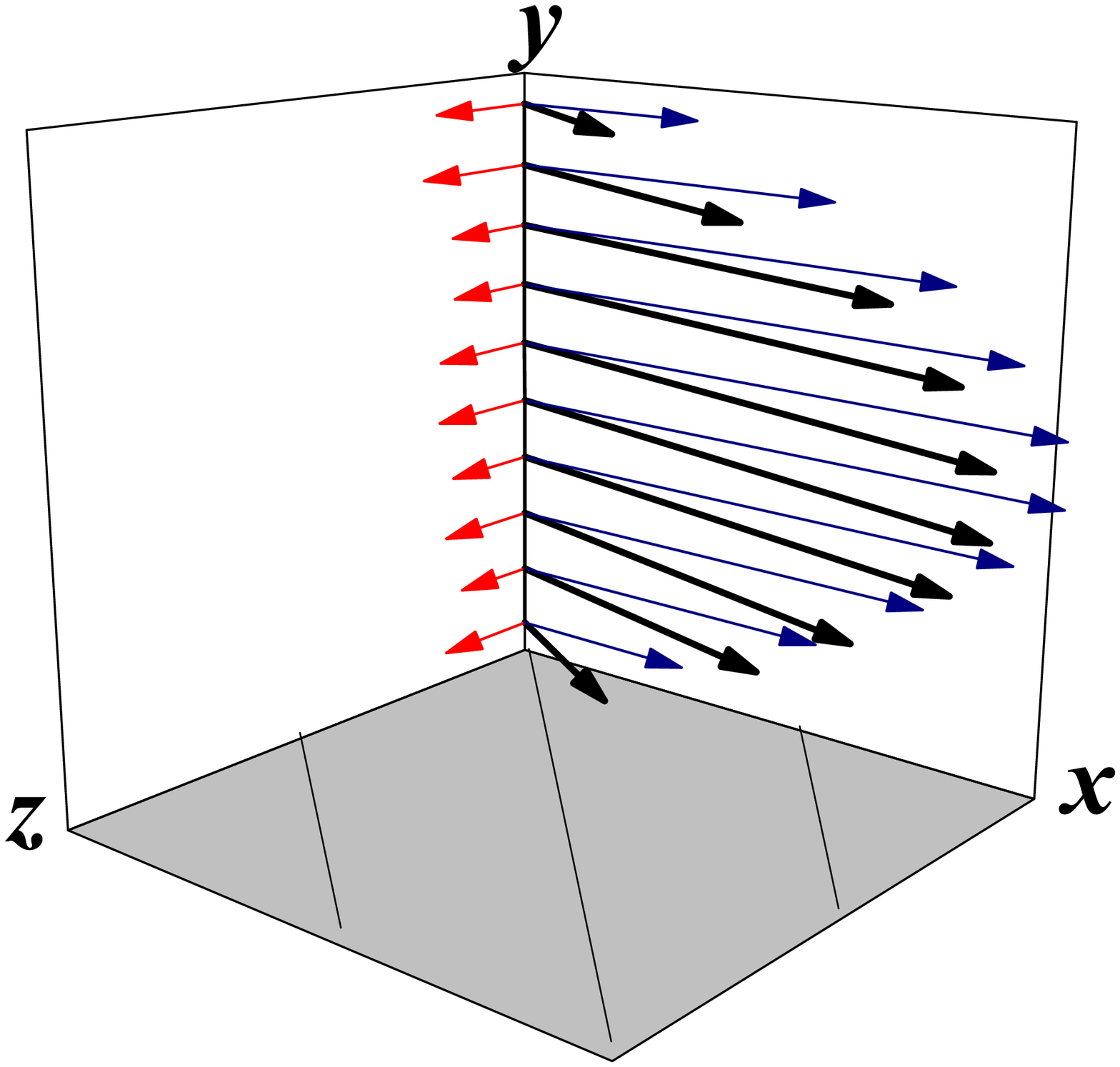}
  (c)\includegraphics[width=0.45\columnwidth]{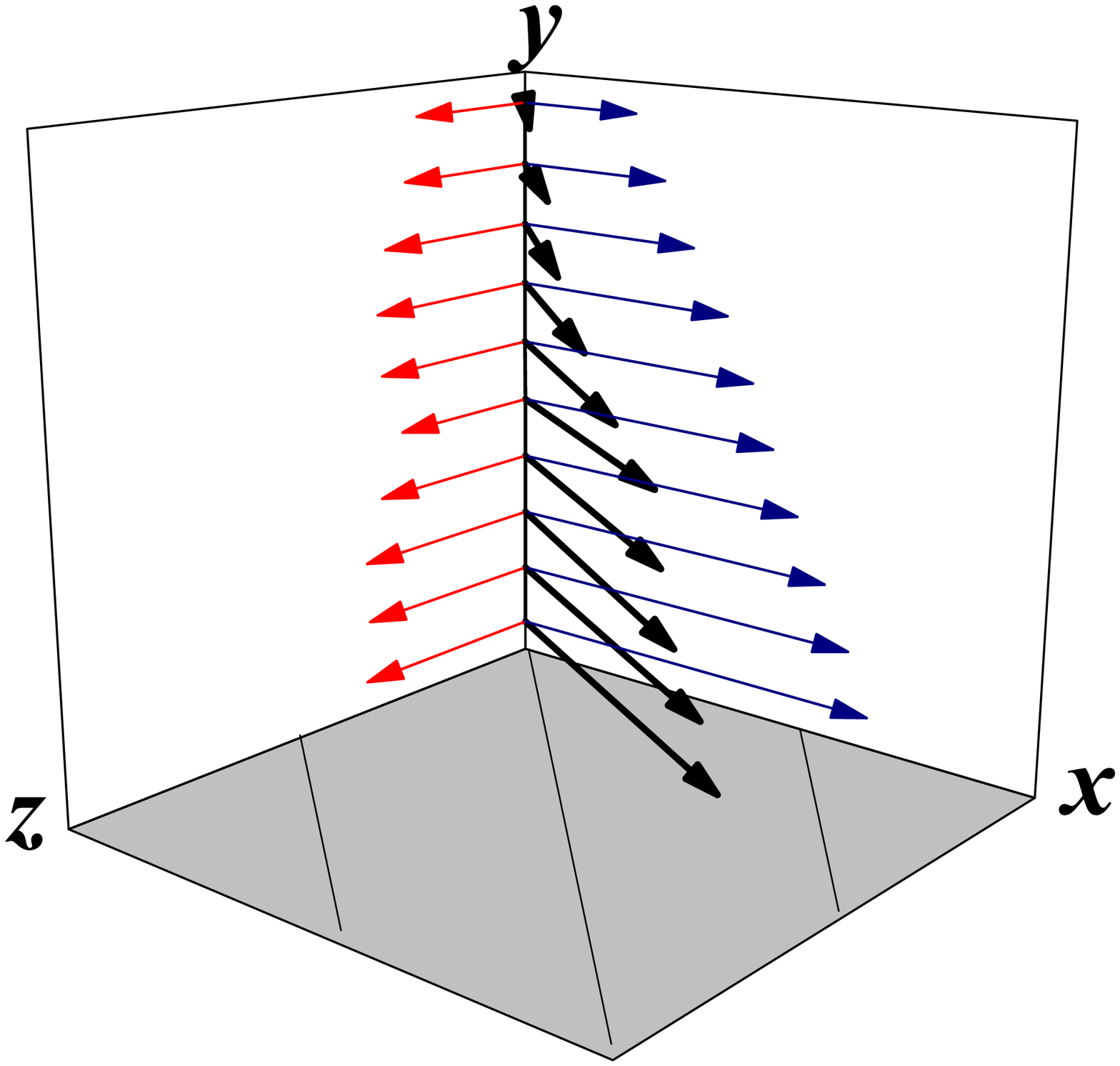}
  (d)\includegraphics[width=0.45\columnwidth]{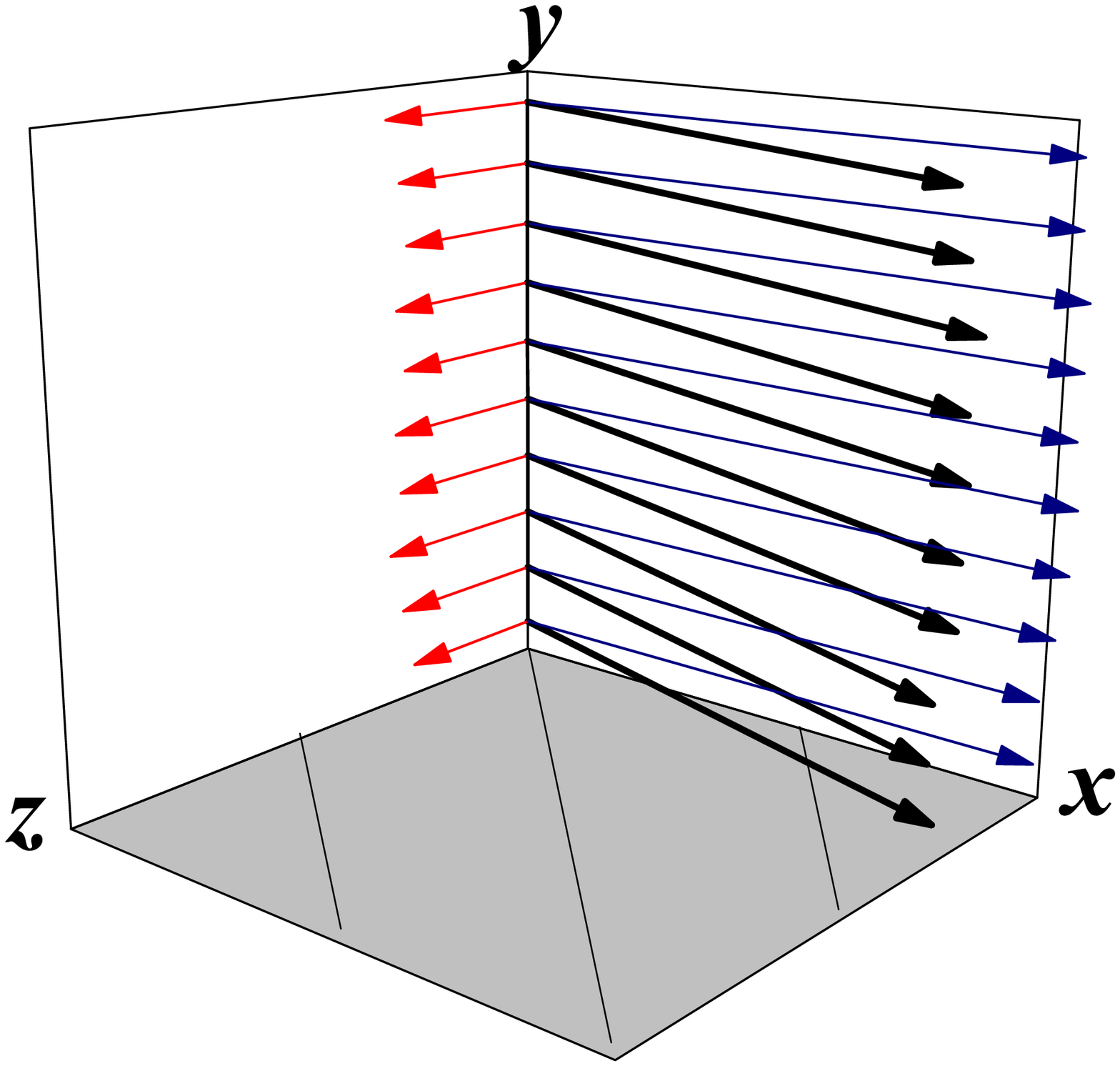}
  \caption{The velocity profile across the channel at $\Theta =\pi/4$: (a)
$H=10\sigma$, $y$ from $-5\sigma$ to $5\sigma$, (b) $H=50\sigma$, $y$ from
$-25\sigma$ to $25\sigma$, (c) $H=50\sigma$, $y$ from $15\sigma$ to $25\sigma$,
enlarged part near the striped-pattern, (d) $H=50\sigma$, $y$ from $-5\sigma$
to $5\sigma$, enlarged part near the channel center. The $z$-components in (c)
and (d) have been increased five times for better demonstration.}
  \label{fig:velocity_prof}
\end{figure*}

Finally in this section, we show the measured velocity profile in a channel
with stripes inclined at the angle $\Theta=\pi/4$ with respect to the applied
force (Fig.\ \ref{fig:velocity_prof}). The velocity profile results from
superimposed flows in the $x$ (forward) and the $z$ (transverse) directions
(also shown). As above we use $H=10\sigma$ [Fig.\ \ref{fig:velocity_prof}(a)]
and $H=50\sigma$ [Fig.\ \ref{fig:velocity_prof}(b)]. The data show that the
effective velocity direction is generally misaligned with the force vector, and
this effect is much more pronounced for a thin channel due to a larger
transverse component of the velocity. The flow has the form of a superimposed
no-slip parabolic Poiseuille flow and a slip-driven plug-flow everywhere in the
thin channel. In case of a thick channel the flow near the surface  is
different from that in the center of the channel. To examine this more closely,
the short-distance region of the flow from Fig.\ \ref{fig:velocity_prof}(b) is
reproduced in Fig.\ \ref{fig:velocity_prof}(c), and the central part is shown
in Fig.\ \ref{fig:velocity_prof}(d). These blowups demonstrate that the strong
transverse flow due to surface anisotropy is generated only in the vicinity of
the wall and tends to disappear far from it, as predicted
theoretically~\cite{vinogradova.oi:2010} and observed in
experiment~\cite{ou.j:2007}. The effective velocity profile is `twisted' close
to the striped wall.

\section{Conclusion}

We have investigated pressure-driven flow in a flat-parallel channel with two aligned striped SH surfaces. For this geometry of configuration we propose a semi-analytical theory, which is valid for an arbitrary gap and any local slip length at the slipping areas of the stripes. Analytical
results are presented for various important limits, and our approach gives  simple analytical formula for
an effective slip length in case of stripes that are inclined with respect to a pressure gradient.
Furthermore, our theory gives analytical guidance as to how to choose the parameters of the texture and a tilt angle, in order to optimize the transverse flow in different situations. Our theoretical predictions have been compared with results of DPD simulations, which are in a good quantitative agreement with the macroscopic theory.
 Our results are directly relevant for passive mixing in thin channels and other microfluidic applications. In the future, we hope to augment our analysis to include the possibility of a
surface charge patterns, which would help to understand transverse electro-osmotic phenomena. Another fruitful direction could be to consider identical, but misaligned striped walls, where the relative orientation of surfaces could generate additional mechanism of mixing.

\section*{Acknowledgments}

We are grateful to V.~Lobaskin for discussions and advice. This research was
supported by the RAS through its priority program `Assembly and
Investigation of Macromolecular Structures of New Generations', and by the DFG through SFB-TR6. The simulations were carried out using computational resources at the John von Neumann Institute for Computing
(NIC J\"ulich), the High Performance Computing Center Stuttgart (HLRS) and Mainz
University.


\end{document}